\documentclass[aps,prb,twocolumn,superscriptaddress,longbibliography]{revtex4-1}
\usepackage[colorlinks=true,citecolor=blue,linkcolor=blue,breaklinks=true]{hyperref}

\usepackage{color,graphicx}
\usepackage{bm}
\usepackage{amsmath}
\usepackage{amssymb}
\usepackage{times}

\allowdisplaybreaks

\newcommand {\up} {\ensuremath{\uparrow}}
\newcommand {\dn} {\ensuremath{\downarrow}}

\newcommand{\lt} {\left}
\newcommand{\rt} {\right}

\begin{document}

\title{The nature of the ``pseudogap" in the insulating phase of highly disordered superconductors}
\author{Mark Nikolaevsky} 
\thanks{These authors contributed equally to this work.}
\address{Department of Physics and Jack and Pearl Resnick Institute and Institute of Nanotechnology and Advanced Materials, Bar-Ilan University, Ramat-Gan 52900, Israel}
\author{Abhisek Samanta}
\thanks{These authors contributed equally to this work.}
\address{Department of Physics, Indian Institute of Technology Gandhinagar, Gujarat 382355, India}
\address{Department of Physics, The Ohio State University, Columbus, Ohio 43210, USA}
\author{Nandini Trivedi}
\address{Department of Physics, The Ohio State University, Columbus, Ohio 43210, USA}
\author{Aviad Frydman}
\address{Department of Physics and Jack and Pearl Resnick Institute and Institute of Nanotechnology and Advanced Materials, Bar-Ilan University, Ramat-Gan 52900, Israel}

\date{\today}

\begin{abstract}
\noindent
Disordered thin films undergoing a superconductor–insulator transition provide a controlled setting for studying pseudogap physics in the absence of competing electronic orders. Although theory predicts that local Cooper pairing can survive deep into the insulating phase, direct spectroscopic confirmation has remained experimentally inaccessible because tunneling measurements in highly insulating films require ultra-high-resistance junctions and picoampere current sensitivity. Here we combine ultra-high-resistance planar tunneling spectroscopy on amorphous indium oxide films with quantum Monte Carlo simulations of the  attractive Hubbard model to probe the single-particle excitation spectrum deep in the insulating regime. We find striking agreement between experiment and theory: the single-particle gap not only survives across the superconductor–insulator transition, but increases substantially with disorder, reaching values more than twice those observed on the superconducting side. With increasing temperature, the gap fills rather than closes, while coherence peaks are suppressed and spectral weight redistributes to energies far exceeding the gap scale. Our results provide direct quantitative experimental confirmation of the theoretically predicted Cooper-pair insulating state with localized Cooper pairs and establish a unified connection between the pseudogap above $T_c$ and the insulating gap as manifestations of pairing without global phase coherence.

\end{abstract}

\maketitle

\noindent {\it Introduction:}
The discovery of high-temperature superconductivity in cuprates in 1987 overturned long-held paradigms: superconductivity was no longer confined to good metals, could be found in ceramics, could coexist with magnetism, and could deviate from BCS behavior through the emergence of a pseudogap~\cite{KeimerKivelson2015,anderson2004physics,lahoud2014emergence,trivedi1995deviations,randeria1998pairing}. In cuprates, the suppression of low-energy spectral weight above $T_c$ has been attributed to a wide range of competing mechanisms arising from strong correlation effects associated with the nearby Mott insulating state. It is argued that these competing mechanisms can lead to fluctuating preformed Cooper pairs, magnetic order, charge order, and orbital order that can generate normal states with a suppressed density of states. A major open question is whether a pseudogap can emerge purely from superconducting pairing without global phase coherence. 

The disorder-driven superconductor–insulator transition (SIT) in homogeneously disordered thin films \cite{strongin1970kammerer, dynes1986breakdown,haviland1989onset, valles1992electron, frydman2002universal, aubin2006magnetic, stewart2007superconducting, sacepe2008disorder, marrache2008thickness, hollen2011cooper, postolova2017reentrant, baturina2011nanopattern,poran2017quantum,shahar1992superconductivity,sacepe2011localization,poran2011disorder,roy2018quantum,paalanen1992low,yazdani1995superconducting,gantmakher1998destruction,sambandamurthy2004superconductivity,sambandamurthy2005experimental,steiner2005possible,baturina2005quantum,baturina2007quantum,crane2007fluctuations,vinokur2008superinsulator,ganguly2017magnetic,mondal2011phase,parendo2005electrostatic,caviglia2008electric} provides a uniquely controlled platform to address this question. In these systems, disorder suppresses long-range phase coherence while preserving local pairing correlations, producing a pseudogap state without the competing orders present in cuprates. Previous tunneling experiments \cite{sacepe2011localization,sherman2012measurement,sherman2014effect} and theoretical studies \cite{bouadim2011single} established that a spectral gap survives near the SIT even after superconductivity is destroyed, consistent with localized Cooper pairs and strong phase fluctuations. However, the fate of this gap deep in the insulating phase has remained unresolved.

The main obstacle has been experimental. Probing tunneling spectra in highly insulating films requires tunnel junction resistances far exceeding the sample resistance in order to maintain an equipotential tunneling electrode. Deep in the insulating regime this necessitates junction resistances on the order of $100M\Omega$, restricting subgap measurements to currents of only a few picoamperes at millivolt bias. Under these conditions, conventional scanning tunneling microscopy becomes impractical because the exponentially suppressed tunnel current cannot reliably stabilize the feedback loop that maintains the tip–sample separation, often leading to loss of tunneling conditions or tip crash. These constraints have prevented reliable spectroscopy far beyond the superconductor–insulator transition for more than a decade. 

\begin{figure*}
\centering  \includegraphics[scale=0.5]{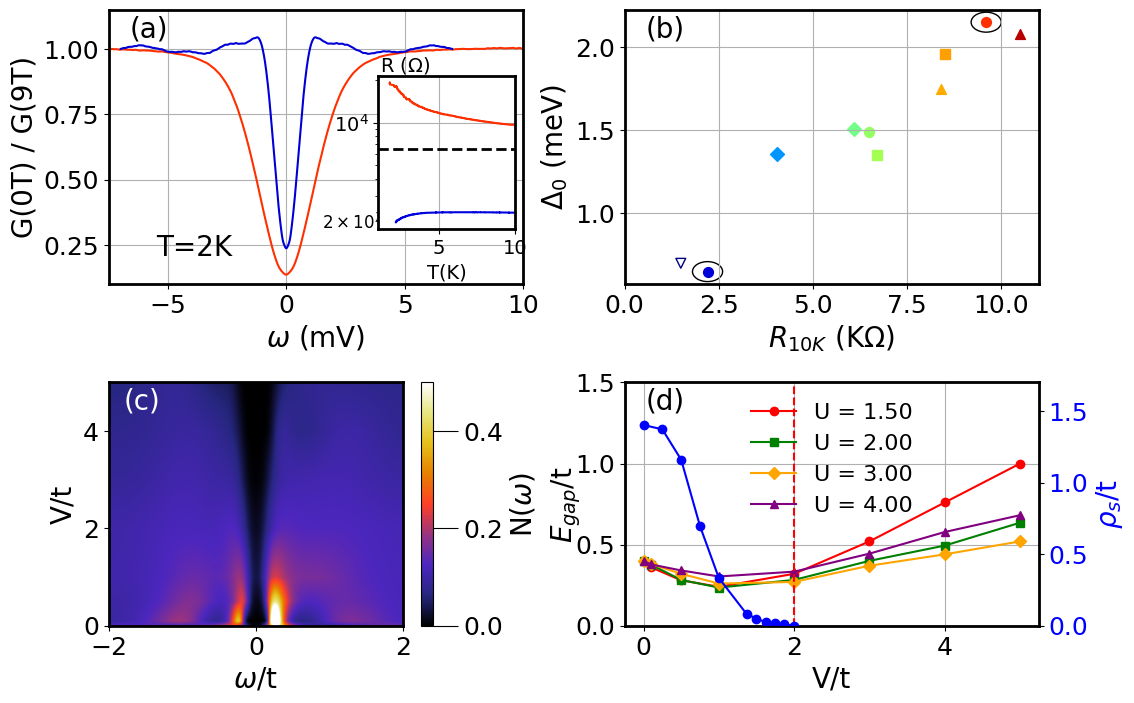}
\caption{\textbf{Effect of the disorder on the gap}. Panel (a) depicts the normalized tunneling density ($G$) of two tunneling spectra from the same sample at two different disorder levels at 2K. The highly disordered measurement is shown in red and a less disordered measurement, on the verge of SIT, is shown in blue. The inset depicts the corresponding resistance $R$ as a function of temperature in red and blue lines respectively. The dashed line shows the quantum of resistance $h/(2e)^2 =6.45 K\Omega$. Panel (b) summarizes the fitted gap as a function of disorder (represented as the resistivity of the sample at 10K) from the experimental results using the Dentelski \textit{et al}. model~[\onlinecite{dentelski2018tunneling}] at 2K. Each sample is shown in a different color, with multiple points showing different disorder levels of the same sample achieved by annealing. The two circled points are the gaps corresponding to the spectra shown in panel (a). Panel (c) is a three-dimensional map of the calculated density of states from the theoretical model as a function of energy ($\omega$) at varying levels of disorder ($V$) with the attractive interaction strength between electrons $U=2t$. Panel (d) depicts the calculated gap against the disorder parameter $V$ at $T=0$ using quantum Monte Carlo (QMC) simulation for different $U$, and the corresponding superfluid density ($\rho_s$) is plotted for $U=4t$ (scale shown on the right). $U$ is the strength of the attractive interaction between electrons in units of hopping $t$. }
\label{fig:gap_to_disorder}
\end{figure*}

Here we overcome this limitation using ultra-high-resistance planar tunnel junctions on amorphous indium oxide films, whose fixed barrier geometry remains stable independently of the sample conductance. Combining these measurements with determinant quantum Monte Carlo simulations of the disordered attractive Hubbard model, we directly probe the single-particle excitation spectrum deep inside the insulating phase and find striking agreement between experiment and theory.

We show that the single-particle gap not only survives across the SIT, but, counterintuitively, it grows substantially with increasing disorder deep in the insulating phase, reaching values more than twice those on the superconducting side of the transition. At elevated temperatures the gap does not close in a BCS-like manner. Instead, subgap states progressively fill in while coherence peaks are strongly suppressed, indicating that phase fluctuations dominate the spectral evolution. We further observe that spectral weight removed from the gap region is redistributed to energies far exceeding the gap scale, consistent with strong localization effects in the disorder potential landscape. To our knowledge, this is the first direct quantitative experimental confirmation of the theoretically predicted growth of the single-particle gap deep in the insulating regime. Together with the quantitative agreement between experiment and quantum Monte Carlo simulations, our results establish a unified connection between the pseudogap above $T_c$ and the insulating gap as manifestations of local Cooper pairing without global phase coherence.

\begin{figure*}
\centering
\includegraphics[scale=0.4]{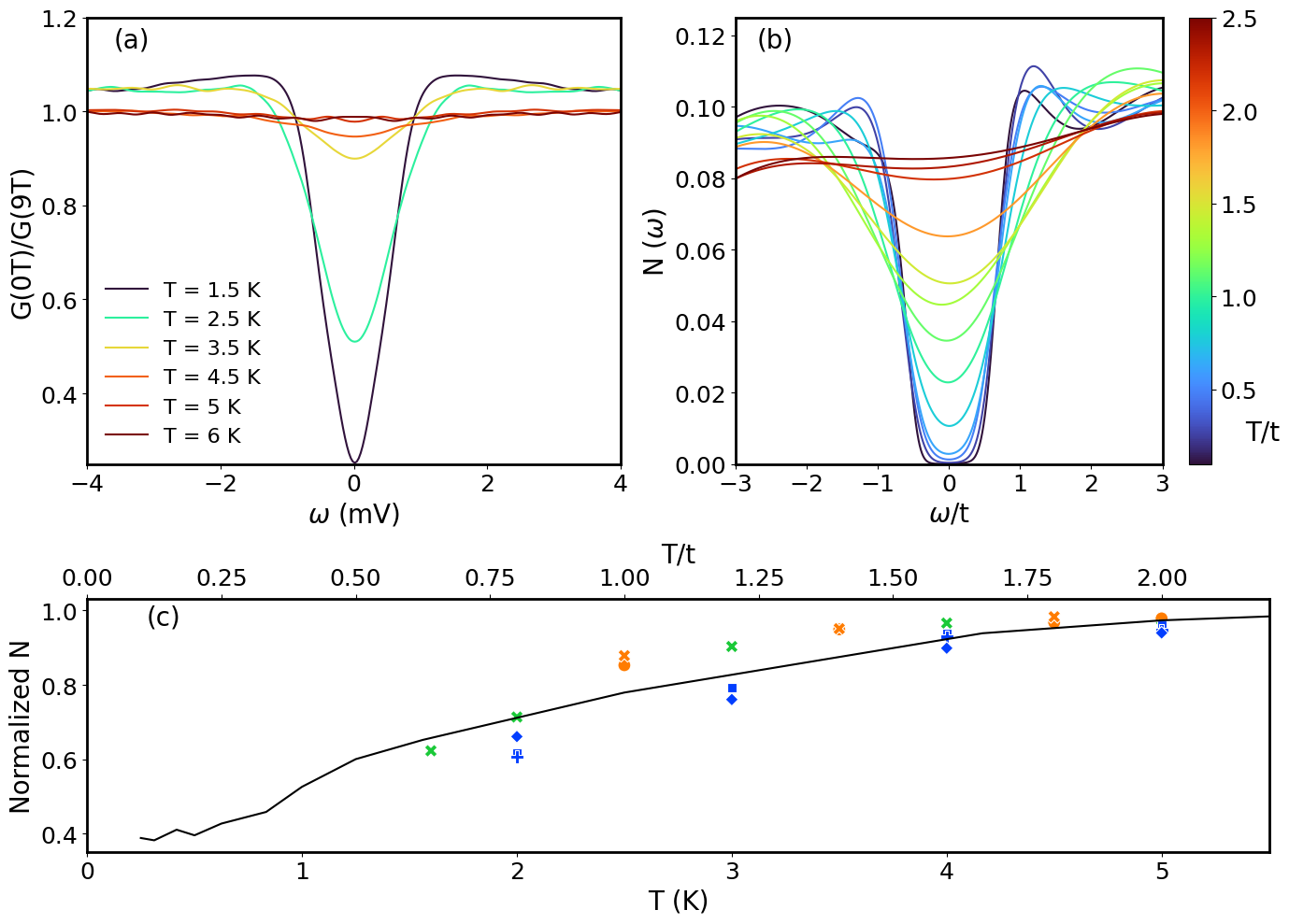}
\caption{\textbf{The density of states (DOS) as a function of temperature}. Panels (a) and (b) show the evolution of the DOS as a function of temperature in the insulating phase from the experimental measurements and the theoretical calculation respectively. (a) depicts the measured normalized tunneling density as a function energy (applied dc voltage) at varying temperatures from 1.5K to 6K, (b) is the calculated DOS as a function of energy ($\omega$) using the DQMC simulations, at a disorder level $V/t=2.5$ and at varying temperatures. Panel (c) shows the temperature dependence of the normalized low-energy DOS, obtained by integrating the spectra within $\pm\Delta$. The symbols represent experimentally extracted values from samples with different disorder levels normalized to their high-temperature values where the gap vanishes, while the solid black curve shows the corresponding DQMC result.}
\label{fig:Thermal_behavior}
\end{figure*}

\noindent {\it Samples:} To measure the energy gap deep in the insulating phase of highly disordered superconductors  we employ planar tunneling junctions with resistances reaching $100M\Omega$. Due to the sub-millivolt voltage biases required to probe the gap, the measured tunneling currents should be in the picoampere range, necessitating ultra-sensitive detection using a designated transimpedance amplifier. The samples studied in this work are thin films of disordered amorphous Indium Oxide ($a-InO$) evaporated in a partial  $O_{2}$ pressure ranging between 2-5$\times10^{-5}$ mbar which produces  highly disordered samples spanning from slightly superconducting to strongly insulating films. For studying the DOS we fabricated $InO/AlO/Al$ planar tunnel junctions. Sample preparation details are described in the Supplementary Material. In order to ensure an equi-potential electrode for tunneling, we require that the resistivity of the junction be at least two orders of magnitude larger than the sample resistance. This sets a limit on the insulator resistance, thus restricting our measurements to fairly high temperatures (above $1.5K$) for strongly disordered films.

\begin{figure*}
\centering
\includegraphics[width=0.8\linewidth]{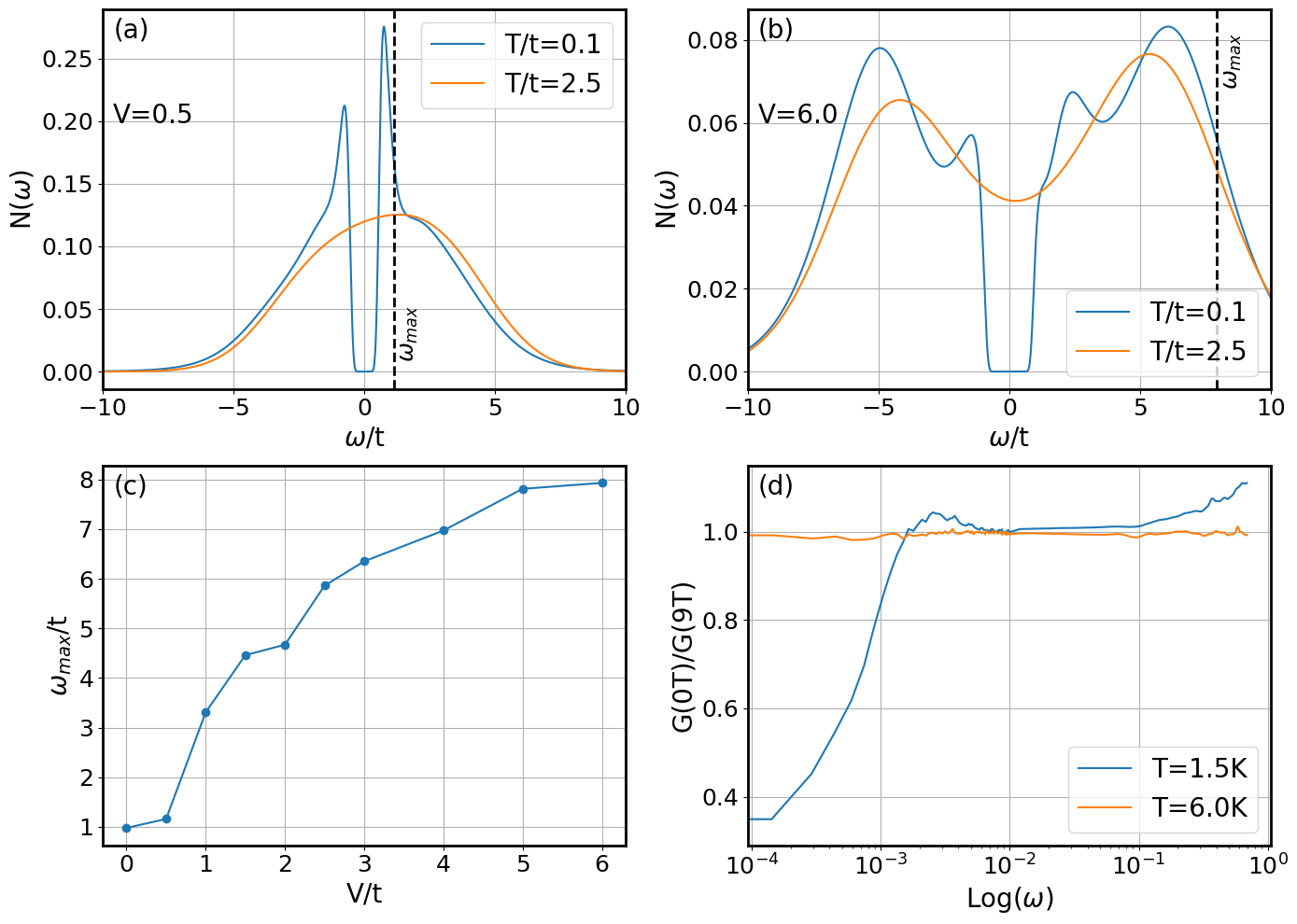}
\caption{\textbf{Suppression of states into higher energies at large disorder}. Panels (a) and (b) depict the calculated density of states (DOS) N($\omega$) based on DQMC simulations at different disorder levels $V=0.5t$ and 6$t$, respectively. For each disorder, N($\omega$) is shown for low temperature (blue line) and a higher temperature (orange line), where the coherence peaks have been suppressed. Note the relative increase in the density of states at high energies at higher temperatures. (c) Maximal energy ($\omega_{\rm max}$) defined by the energy cut-off that equalizes the integrated DOS for the low temperature curve below the cut-off to corrresponding area for the high temperature curve. This gives a quantitative measure for the depletion of states around the coherence energy and an increase at higher energies. The dashed lines in figures (a) and (b) represent $\omega_{\rm max}$ for the corresponding disorder values. (d) Measurement of the normalized tunneling density as a function of energy (applied dc voltage) at two different temperatures, below $T_{c}$ at 1.5K (blue line) and above $T_{c}$ at 6K (orange line) where we see an increase in the density of states with respect to the energy at higher temperatures.}
  \label{fig:high_energy}
\end{figure*}

\bigskip

\noindent {\it The gap at high disorder:} 
Fig.~\ref{fig:gap_to_disorder} (a) compares  the tunneling conductance $G=dI/dV$ versus $V$ curves, normalized to the normal state curves at $H=9$T, on two insulating samples, one highly disordered and the other close to the SIT. The high disorder film clearly exhibits a larger energy gap  $\Delta$. This trend was found for nine measured films as shown in Fig.~\ref{fig:gap_to_disorder} (b) that depicts $\Delta$ as a function of the sheet resistance at 10K, $R_{\square}$, which is taken as our experimental measure of the disorder. Here $\Delta$ is evaluated using a model that takes into account the fluctuation in space and time of the order parameter in the disordered film \cite{dentelski2018tunneling} (see supplementary material) for details). It is seen that $\Delta$ monotonically increases with disorder, reaching a value which is more than a factor of two larger than that of low disorder. 

 \begin{figure}
\centering 
\includegraphics[width=0.99\columnwidth]{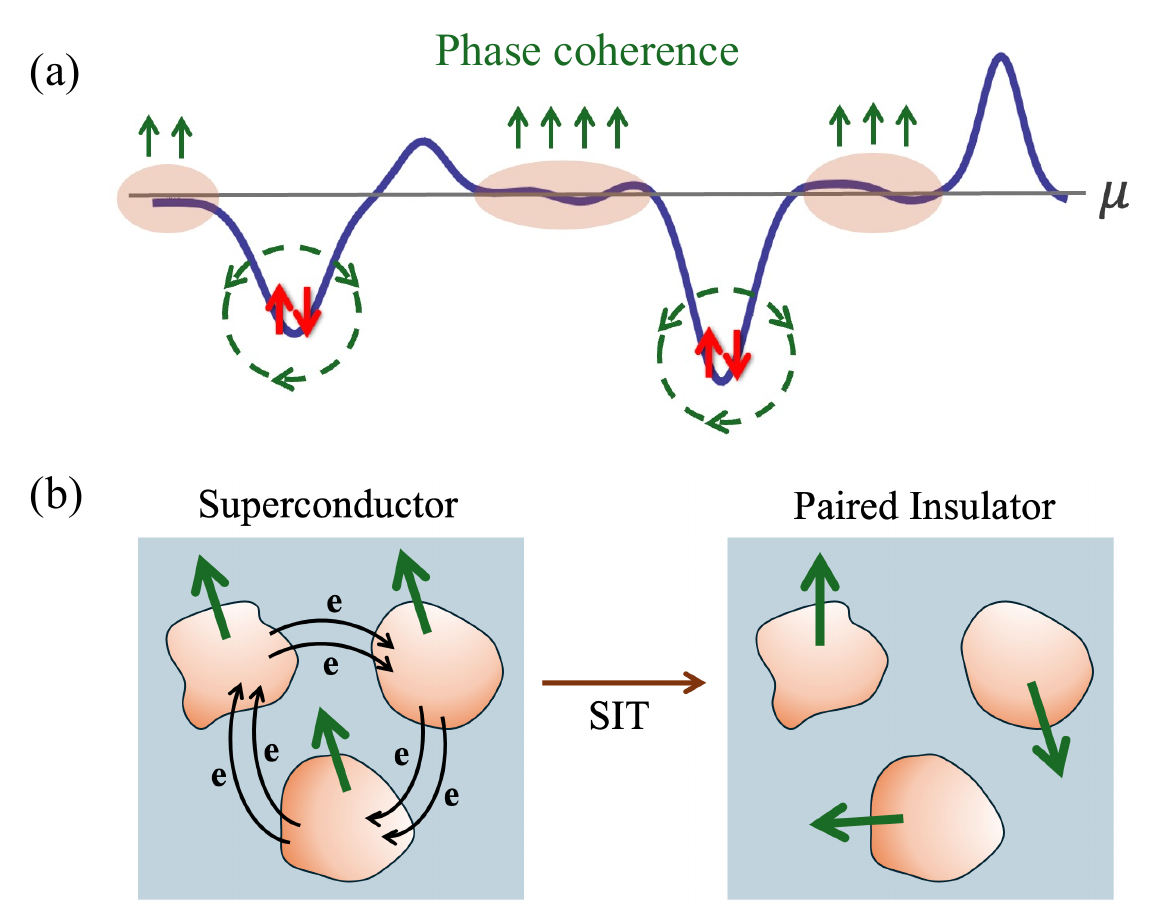}
\caption{Schematic illustration of superconductivity in a highly disordered potential in the insulating phase of the superconductor–insulator transition (SIT). The gray surface represents a spatially random disorder potential, with red spheres denoting bound electron pairs localized in potential minima. Purple regions indicate locally coherent superconducting (SC) islands, with arrows representing the local SC phase within each island. At strong disorder, large potential barriers suppress coupling between islands, preventing global phase coherence across the sample.}
\label{fig:illustration}
\end{figure}

Our theoretical analysis shows a similar trend. We consider an attractive Hubbard model on a square lattice in the presence of non-magnetic onsite disorder. The attractive interaction is responsible for the formation of Cooper pairs. The model Hamiltonian is given by,
\begin{align}
\label{eq:H}
H &= -t\!\!\sum_{\langle ij\rangle,\sigma}\!\! \lt(c^\dagger_{i\sigma}c_{j\sigma} +H.c.\rt) \!+\!\sum_i(V_i\!-\!\mu)n_i
\!-\! U\!\sum_i n_{i\up}n_{i\dn}
\end{align}
where $t$ is the hopping amplitude of an electron between two nearest-neighbor sites, $U$ is the strength of the attractive interaction between two electrons, and $\mu$ is the chemical potential which tunes the density of electrons. $V_i$ is a random number drawn uniformly from a distribution of mean 0 and width $2V$ i.e. $V_i\in[-V,V]$, where $V$ sets the strength of the disorder in our problem. We study this model by using determinantal quantum Monte Carlo (DQMC) simulations, of which the details are provided in the supplementary material. Importantly, this method includes the amplitude and phase fluctuations arising from both thermal and quantum origins (see supplementary material for details).
Fig.~\ref{fig:gap_to_disorder} (c) shows the density of states (DOS) $N(\omega)$ obtained from DQMC simulations which affirm the experimental trend.
In particular, the gap grows significantly with increasing disorder and may reach more than a factor of 2 increase at large disorder for a weak electron coupling, $U=1.5t$ (Fig.~\ref{fig:gap_to_disorder} (d)). In addition, the gap shows a non-monotonic behavior with disorder near the SIT -- below the transition it corresponds to a superconducting gap, while above the transition the gap occurs in a paired insulating state.

The single particle energy gap in the insulating state can become significantly larger than in the superconducting state, as predicted~\cite{bouadim2011single}, due to electron localization on the length scale $\xi$ set by disorder.
This localization enhances the repulsion energy experienced by Cooper pairs, effectively increasing the gap $\Delta \propto U/2\xi^2$. Both our experimental measurements and DQMC simulations show that this trend extends deep into the insulting regime significantly increasing the energy gap.

\bigskip

\noindent {\it Effect of temperature at high levels of disorder:} 
The temperature dependence of the density of states (DOS), shown in Fig.~\ref{fig:Thermal_behavior}, is extended in this work deep into the insulating regime of highly disordered InO. Experimentally we find (Fig.~\ref{fig:Thermal_behavior} (a)) that increasing the temperature does not reduce the gap energy in the insulating InO. Instead, $\Delta$ remains constant up to a temperature $T^*$ at which point the gap vanishes. Hence, rather than ``closing" $\Delta$, the temperature causes a filling up of the subgap states until, eventually,  the gap is entirely suppressed. This temperature evolution is similar to that observed at lower disorder, where the sample remains superconducting. 

A similar trend of the thermal behaviour is obtained in our theoretical calculations. Fig.~\ref{fig:Thermal_behavior} (b) depicts DOS versus temperature obtained from DQMC calculations. 
showing that the  coherence peaks (the pile-up of states at the gap edge) remain almost at the same energy ($\omega\sim E_0$) while the gap fills up with temperature, consistent with the experimental results. 
A  quantitative comparison of the gap filling of the experimental and theoretical results is presented in Fig.~\ref{fig:Thermal_behavior} (c)  that shows a measure for the low energy states filling.
 
We note that this temperature dependence, in which the gap fills, rather than closes as expected within BCS theory, also occurs in cuprates\cite{tromp2023puddle} and is considered as a fingerprint of strong coupling $d$-wave superconductivity. In the highly disordered $s$-wave superconductors studied here we show that very similar behavior is found even in a weak coupling superconductor that is nevertheless driven out of the BCS paradigm by {\em disorder} realizing effectively the behavior akin to a strongly coupled superconductor.

\bigskip

\noindent {\it Redistribution of spectral weights with disorder:} Figures \ref{fig:gap_to_disorder} and \ref{fig:Thermal_behavior} demonstrate that, on the insulating side of the SIT, the gap size remains large, and even grows with increasing disorder, however the coherence peaks vanish. This behavior is attributed to the lack of coherence in the sample as the superconducting regions decouple, while local superconducting gap remains finite with a fluctuating phase. This poses a question as to regarding the conservation of spectral density. What compensates for the missing states in the gap?

We find that these  missing states are pushed to energies much higher than $\Delta$ as shown in the in Fig.~\ref{fig:high_energy} (a)-(c), where the calculated density of states is shown at different levels of disorder. This finding is confirmed experimentally in Fig.~\ref{fig:high_energy} (d) which compares the measured $G$ as a function of $\omega$ for $1.5K$ and $6K$ where the gap is suppressed. A steady increase in the density of states is measured at high energies. Although the theory and experiment agree qualitatively on the high-energy redistribution of spectral weight, the energy scale differs significantly. In the calculations, the shift is limited to the same order of magnitude as $\Delta$, whereas experimentally it extends nearly two orders of magnitude. We attribute this discrepancy to the finite bandwidth inherent to the numerical model. In the calculations, the shift is limited by the finite numerical bandwidth, whereas experimentally the spectrum is effectively unbounded, permitting redistribution over a much wider energy range.

The behavior highlighted above can be understood qualitatively in the following way: given a certain high-disorder profile for a film, we can broadly describe the topography as consisting of plateaus, hills and valleys. Upon overlaying the pairing amplitude map $\Delta({\bf r})$ on this disorder landscape, we notice the following: 

\noindent (a) The pairing amplitude is large in the plateau regions that allow for pairing of electrons and particle-hole fluctuations about the chemical potential. These regions form locally coherent superconducting islands, within which phase coherence can build up over a length scale comparable to the coherence length. 

\noindent (b) The pairing amplitude is zero or small in the deep valleys, which host bound pairs of opposite-spin electrons; yet $\Delta({\bf r})$ remains small due to the absence of any particle–hole mixing. These localized pairs lead to bound states trapped in potential minima in the insulating phase (Fig.~\ref{fig:illustration} (a)). 

\noindent (c) The pairing amplitude is also zero or small in the hills since there is no electron density in these regions. As disorder increases, the superconducting islands on the plateaus become increasingly isolated by large potential barriers (hills and deep valleys), suppressing inter-island Josephson coupling. Although local pairing and phase coherence persist within individual islands, global phase coherence across the sample is destroyed, driving the system into a paired insulating state via a superconductor--insulator transition (SIT) (Fig.~\ref{fig:illustration} (b)). Consistent with this picture, the density of states in the insulating phase shows a redistribution of spectral weight to higher energies (Fig.~\ref{fig:high_energy}) over a scale set by the disorder strength -- an energy range several times the gap -- reflecting the disorder scale of the valleys at negative bias and the hills for positive bias.

\noindent {\it {Summary:}}
We experimentally probed the single-particle excitation spectrum deep in the insulating phase of a disorder-driven superconductor–insulator transition using ultra-high-resistance planar tunneling spectroscopy. We find that the insulating pseudogap not only survives beyond the transition but grows substantially with increasing disorder, while filling rather than closing with temperature. The strong agreement between experiment and theory provides direct evidence for a Cooper-pair insulating state governed by local pairing without global phase coherence.  More broadly, our results establish a close connection between disorder-induced pseudogap physics and phenomenology observed in high-$T_c$ superconductors, suggesting that local pairing correlations alone can generate pseudogap behavior even in the absence of competing electronic orders. An important future direction is to investigate how magnetic-field-induced vortices evolve across the transition from weakly disordered superconductors to the Cooper-pair insulating regime.

\noindent {\it {Acknowledgments:}} This research was supported by grant no 2020331 from the United States -- Israel Binational Science Foundation (BSF).

\clearpage
\appendix

\section{Experimental Details}

The tunneling sample preparation and experimental methodology closely followed a technique outlined in the referenced papers \cite{sherman2012measurement,sherman2014effect}. The experiment involved applying varying dc voltage and a substantially smaller ac voltage ($0.1mV$) to the tunneling junction. The dc voltage was scanned over the range of $-100mV$ to $100mV$, with the ac current measured using an ac current-to-voltage amplifier (FEmto DDPCA-300).

From the measured ac current, the derivative of the current with respect to the ac voltage ($\frac{dI}{dV}$), or junction conductivity, was calculated as a function of the varying dc voltage, yielding the tunneling spectrum.

\begin{figure}
\centering 
\includegraphics[width=0.98\columnwidth]{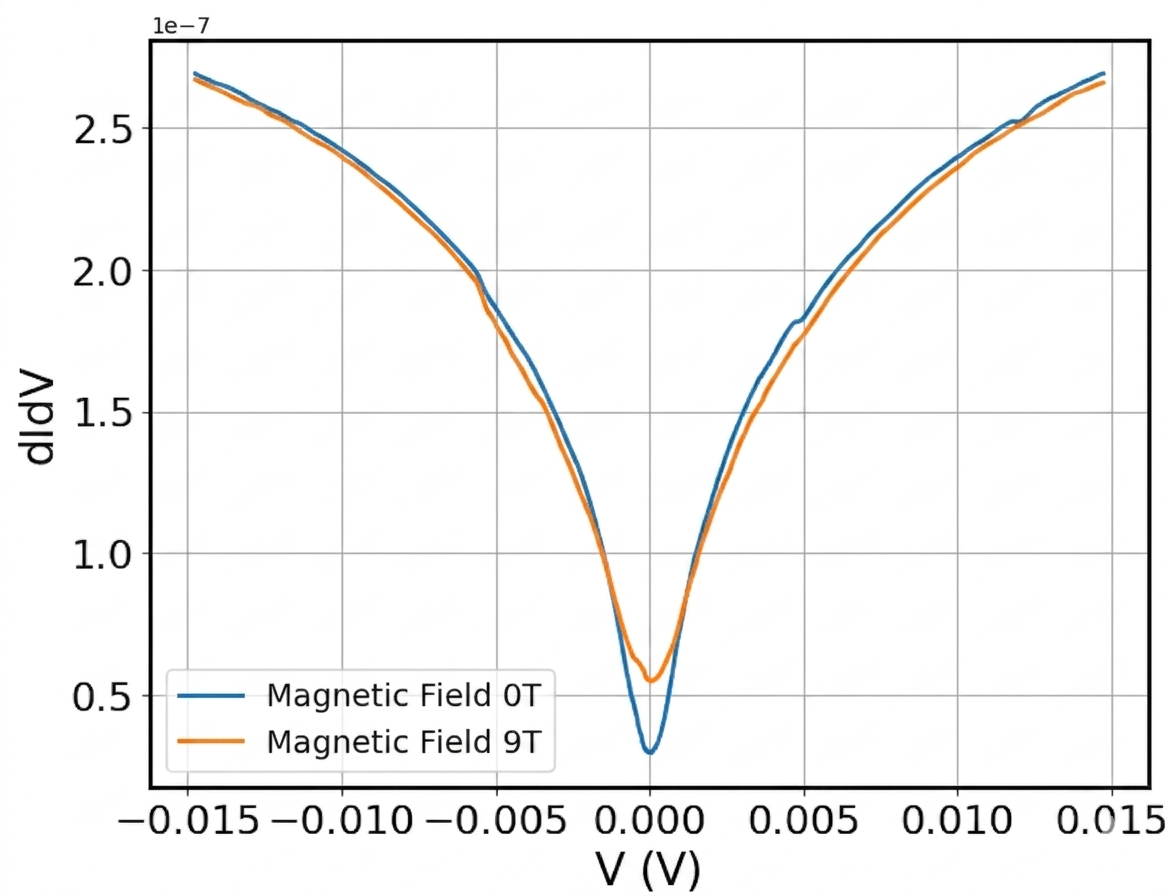}
\caption{{\bf An example of an experimental measurement before analysis.} $dI/dV$ measurement as a function of Volts of one of the samples at 2.5K. The plot two measurements taken at H=0T and H=9T. The final normalized $dI/dV$ result is obtained by dividing the two plots, this is done to remove effects that are persistent at high magnetic fields.}
\label{fig:raw}
\end{figure}

The measurements were conducted at temperatures as low as $1.5K$ and magnetic fields as high as $9T$ using a He4 cryostat produced by attoCube. The tunneling spectra were collected at different temperatures ranging from $1.5K$ to $10K$. At each temperature, the spectrum was taken at both $0T$ and $9T$. An example of a resulting measurement can be seen in  Fig. ~\ref{fig:raw}. 

\begin{figure}
\vspace{0.5cm}
\centering
\includegraphics[width=0.98\columnwidth]{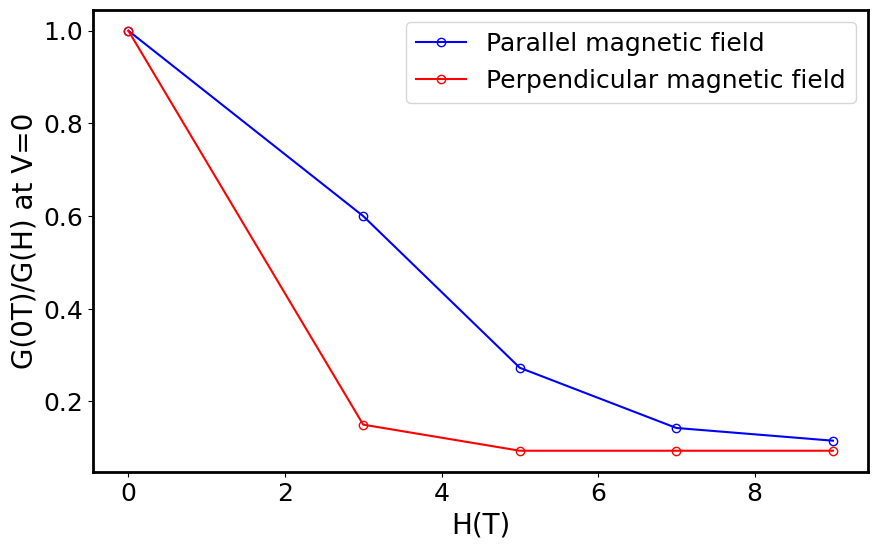}
\caption{{\bf Comparison between the effect of parallel vs. perpendicular magnetic field on the samples.} Zero-bias conductance at $H=0$, normalized by its value at field $H$, for the same sample measured with magnetic field applied parallel (blue) and perpendicular (red) to the film. A perpendicular field can introduce vortices, which locally create subgap states and would therefore enhance the zero-bias conductance. Although the low-field responses differ, both orientations converge to essentially the same value at $H=9T$. This demonstrates that vortex-induced subgap states cannot account for the observed filling of the gap. }
\label{fig:perp vs par}
\end{figure}

To comprehensively assess the magnetic contribution to the measurement, some samples underwent measurements in both perpendicular and parallel magnetic fields. This approach aimed to test the possibility that the filling of states below the energy of the gap was induced by vortices. The results for this are shown in Fig. ~\ref{fig:perp vs par}.

The tunneling spectrum was then employed to evaluate the superconducting energy gap by fitting it to a model published by David et al. in 2018, offering a quantitative understanding of the gap characteristics. Moreover, the spectrum was used to study the disappearance of coherence peaks, providing insights into the behavior of coherence peaks within the insulating state of $InO$.

\begin{figure}
\centering 
\includegraphics[width=0.6\columnwidth]{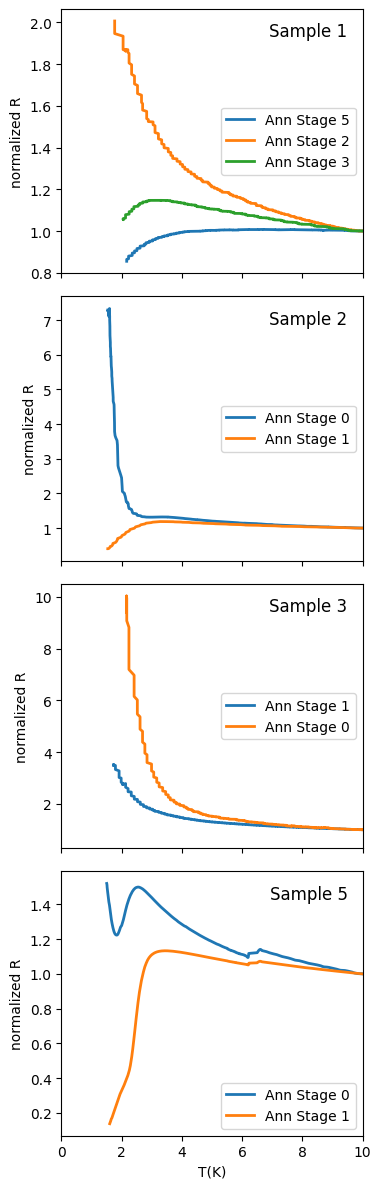}
\caption{{\bf Resistance as a function of temperature.} The resistance as a function of temperature for all the measured successful samples. each plot shows a different sample with varying annealing stages representing the disorder of the sample (The highly disordered samples are reduced by annealing).}
\label{fig:Sup_RT}
\end{figure}

Four samples were measured, all of them evaporated in high oxygen pressure, initiating the measurement in the insulating phase. Most of the samples underwent multiple measurements. Between each measurement, the samples were annealed at $80^\circ\mathrm{C}$ to reduce disorder, transitioning from the insulating regime toward the SIT critical point. The Resistance as a function of temperature of all the successful sample are plotted in fig.~\ref{fig:Sup_RT}.

\section{Details of Determinant Quantum Monte Carlo (DQMC) Simulations}

\begin{figure*}
\centering 
\includegraphics[width=1.7\columnwidth]{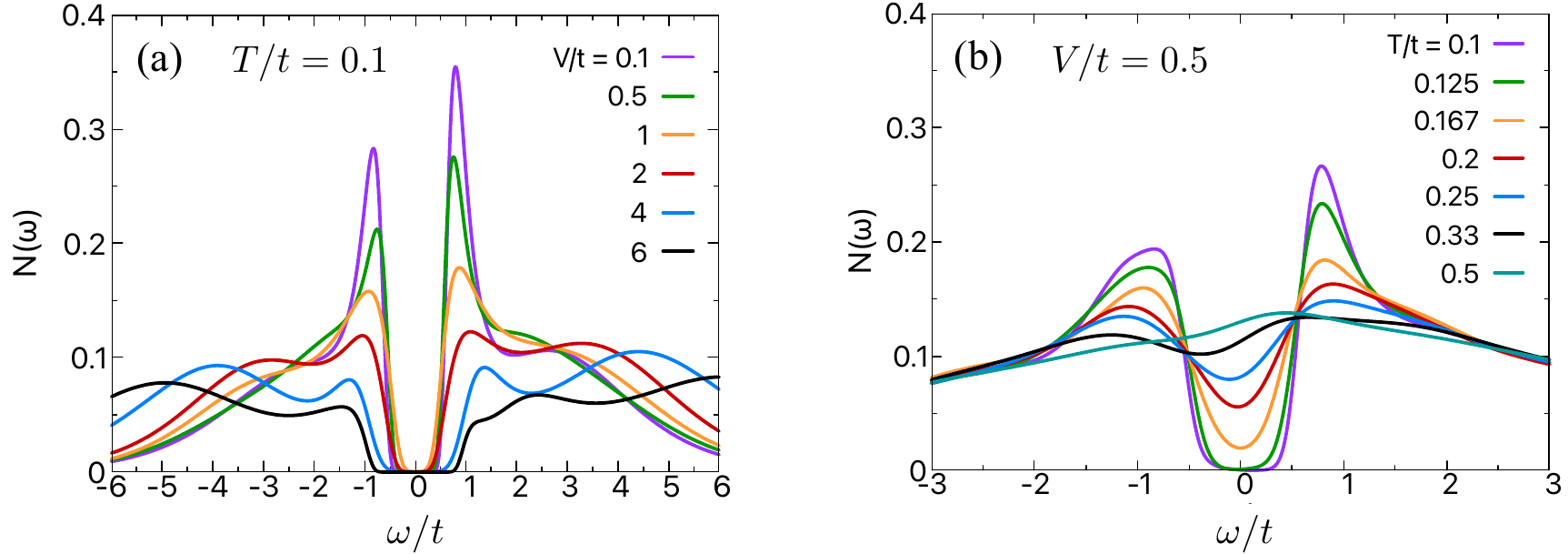}
\caption{{\bf Effects of quantum and thermal fluctuations.} The density of states $N(\omega)$ at $U/t=4$ for (a) fixed $T/t=0.1$ and with increasing disorder $V$, and (b) fixed temperature $V/t=0.5$ and with increasing temperature $T$. (a) shows the effects of strong quantum fluctuations and (b) shows the effects of strong thermal fluctuations. The QMC simulations are able to capture the effects of both the fluctuations.}
\label{fig:fluct}
\end{figure*}

To incorporate the effects of both quantum and thermal fluctuations, we perform quantum Monte Carlo (QMC) simulations of the disordered attractive Hubbard model [Eq.~(\ref{eq:H})]. In particular, we compute the imaginary-time local Green's function,
\begin{align}
G({\bf r},\tau) = \mathcal{T}\langle c_{{\bf r}\sigma}(\tau)c^\dagger_{{\bf r}\sigma}(0)\rangle.
\end{align}
As the system approaches the weak-coupling limit (i.e., smaller interaction strength $U$), the correlation length increases, necessitating larger system sizes $L$. We therefore consider lattice sizes ranging from $L=8\times8$ to $L=16\times16$ for interaction strengths between $U/t=4$ and $U/t=1.5$. For each disorder strength, the results are further averaged over 20 independent disorder realizations.

Each simulation is equilibrated for up to $10^6$ Monte Carlo steps. The determinant QMC (DQMC) method employed here is free from the sign problem because the attractive ($-U$) interaction makes the partition function as the square of the contribution from each spin. To obtain real-frequency quantities, we perform analytic continuation of the imaginary-time Green's function $G({\bf k},\tau)$ using the maximum entropy method (MEM), yielding the spectral function $A({\bf r},\omega)$ via
\begin{align}
G({\bf r},\tau) = -\int_{-\infty}^{\infty} d\omega\, \frac{e^{-\tau\omega}}{1 + e^{-\beta\omega}} A({\bf r},\omega).
\end{align}
The average density of states (DOS), $N(\omega)$, is then obtained from the analytic continuation of the spatially averaged Green's function, $\sum_{\bf r} G({\bf r},\tau)$. We extract the superconducting gap $\Delta$ by tracking the peak-to-peak separation between the two largest coherence peaks in $N(\omega)$ located around $\omega = 0$.

\section{Effects of Quantum Fluctuations and Thermal Fluctuations}

The superconductor-insulator transition (SIT) is a remarkable example of a quantum phase transition driven by the interplay between phase coherence and localization. In the weak disorder regime, mean-field theory captures the essential physics~\cite{ghosal2001inhomogeneous}, including the formation of ``superconducting puddles" and suppression of superfluid stiffness. However, close to the transition, both quantum and thermal fluctuations play crucial roles in determining the transport and thermodynamic properties. Quantum fluctuations of the superconducting order parameter dominate at zero temperature and can suppress long-range phase coherence even when local pairing amplitude remains finite, leading to a Bose-insulating state. Thermal fluctuations, on the other hand, become dominant at finite temperatures and drive a superconductor to a metallic-like state through the unbinding the Cooper pairs. In two-dimensional thin films, reduced dimensionality enhances both types of fluctuations, making long-range order particularly fragile. The interplay between these quantum and thermal fluctuations thus governs the critical behavior near the SIT -- an effect completely missed by mean-field theory but accurately captured by our quantum Monte Carlo simulations.

In Fig.~\ref{fig:fluct}, we plot the density of states 
$N(\omega)$ for two representative cases: (a) at a low temperature $T/t=0.1$, varying disorder strength 
$V$ across the SIT, and (b) at a low disorder $V/t=0.5$, varying temperature $T$. Fig.~\ref{fig:fluct} (a) highlights the effect of quantum fluctuations, as the temperature is kept sufficiently low. With increasing disorder, the coherence peaks are strongly suppressed, accompanied by a strong redistribution of spectral weight from the coherence peaks to much higher energies. This behavior reflects the growing influence of quantum fluctuations, under which the superconducting gap initially decreases and subsequently increases as the system transitions into the insulating regime. In contrast, Fig.~\ref{fig:fluct} (b) illustrates the effect of thermal fluctuations, as the disorder is held fixed at a low value (a superconducting state). With increasing temperature, the superconducting gap gradually fills in rather than closes, deviating from the conventional BCS expectation -- a non-BCS feature arising from strong thermal fluctuations.

\end{document}